\documentclass[twocolumn,showpacs,preprintnumbers,
prb,superscriptaddress]{revtex4-1}%
\usepackage{amssymb}
\usepackage{amsmath}
\ifx\msidvipdf\msiundefined
\usepackage{graphicx}
\else%
\usepackage[dvipdfm]{graphicx}
\fi
\usepackage{dcolumn}
\usepackage{bm}
\usepackage{amsfonts}%
\usepackage{xcolor}
\providecommand{\U}[1]{\protect\rule{.1in}{.1in}}

\ifx\pdfoutput\relax\let\pdfoutput=\undefined\fi
\newcount\msipdfoutput
\ifx\pdfoutput\undefined\else
\ifcase\pdfoutput\else
\ifx\paperwidth\undefined\else
\ifdim\paperheight=0pt\relax\else\pdfpageheight\paperheight\fi
\ifdim\paperwidth=0pt\relax\else\pdfpagewidth\paperwidth\fi
\fi\fi\fi
\ifx\msidvipdf\msiundefined\else
\AtBeginDvi{}
\fi

\begin{document}
\title{Dynamically screened Coulomb interaction in the parent compounds of hole-doped cuprates, trends and exceptions}
\author{F. Nilsson}
\email{fredrik.nilsson@teorfys.lu.se}
\affiliation{Department of Physics, Division of Mathematical Physics, Lund University, Professorsgatan 1, 223 63 Lund, Sweden}
\author{K. Karlsson}
\affiliation{Department of Engineering Sciences, University of Sk\"{o}vde, SE-541 28
Sk\"{o}vde, Sweden}
\author{F. Aryasetiawan}
\affiliation{Department of Physics, Division of Mathematical Physics, Lund University, Professorsgatan 1, 223 63 Lund, Sweden}

\begin{abstract}
Although the cuprate high-temperature superconductors were discovered already 1986 the origin of the pairing mechanism remains elusive.
While the doped compounds are superconducting with high transition temperatures {\em T}$_{c}$ the undoped compounds are insulating due to the 
strong effective Coulomb interaction between the Cu $3d$ holes.
We investigate the dependence of the maximum superconducting transition temperature, {\it T}$_{c\text{ max}}$, on the onsite effective Coulomb interaction
$U$ using the constrained random-phase approximation. We focus on the commonly used one-band model of the cuprates, including only the 
antibonding combination of the Cu $d_{x^2-y^2}$ and O $p_x$ and $p_y$ orbitals, and find a clear screening dependent trend between the static value of $U$ and $T_{c\text{ max}}$ 
for the parent compounds of a large number of 
hole-doped cuprates. Our results suggest that superconductivity is favored by a large onsite Coulomb repulsion. We analyze both the trend in the static value of $U$ and its frequency dependence in 
detail and, by comparing to other works, speculate on the mechanisms behind the trend.
\end{abstract}

\maketitle

\section{Introduction}
One of the most important experimental insights about the high-temperature
copper oxide superconductors is how $T_{c}$ is correlated with the materials
structure expressed as functions of doping, pressure and compositions. The
phase diagrams of the high $T_{c}$ cuprates as a function of doping
concentration reveal a generic feature common to all compounds, showing the
characteristic parabolic curve separating the superconducting and the normal
phases. The crystal structure of the cuprates exhibit the generic copper-oxide planes
where the dominant low-energy physics is beleived to be constrained.
It is known for a long time that $T_{c\text{ max}}$ increases with the number
of CuO$_{2}$ layers and for a given number of layers there is a strong
dependence of $T_{c\text{ max}}$ on the cuprate family. It was, however, not
known at the microcsopic level on which quantum mechanical parameters
$T_{c\text{ max}}$ depended. This puzzle was investigated and analyzed in
detail by Pavarini \emph{et al.} more than a decade ago and they found an
interesting and important trend showing a correlation between $T_{c\text{
max}}$ and the hopping parameters \cite{Pavarini01Band}.
Thorough investigation of the phase diagrams of the high $T_{c}$ cuprates on
the other hand has revealed that the macroscopic properties of the copper
oxides are decisively influenced by strong electron-electron interaction
(large Hubbard $U$) between the copper $3d$ holes (see, e.g. Ref. \onlinecite{Lee06Doping}).
The large Coulomb repulsion also
profoundly influences other fundamental properties which do not follow the
standard Fermi-liquid theory which is exhibited already in the case of zero
doping in which the material becomes an antiferromagnetic Mott insulator.
While a large Coulomb repulsion is at first thought not conducive for the formation of
Cooper pairs leading to superconductivity,
theoretical studies of the two dimensional single-band Hubbard model indicate
that superconductivity can be favored by a large $U$\cite{Fukuyama87Critical,yokoyama2012crossover,Kaczmarczyk13Superconductivity,yanagisawa2015high,yanagisawa2016superconductivity}.

The Heisenberg nearest neighbor exchange parameter $J$ is a quantity that is intimately related to the Hubbard $U$.
In the large $U$ limit the quantities are directly related as $J=-4t^2/U$, where $t$ is the nearest neighbor hopping.
In a recent experimental study\cite{Mallet13Dielectric} Mallet \emph{et al.}
investigate the dependence of $J$
on $T_{c\text{ max}}$ for the systems R(Ba,Sr)$_2$Cu$_3$O$_y$. 
It was shown that $J$ had a strong correlation with $T_{c\text{ max}}$ for the considered compounds.
However, it was also shown that changing internal pressure by ion-substitution and varying the external pressure have
identical effects on $J$ but opposite effects on $T_{c\text{ max}}$. On the other hand
the refractivity sum was shown to have a strong correlation with $T_{c\text{ max}}$
which lead the authors to suggest a dielectric rather than a magnetic pairing mechanism.

The purpose of the present work is to delve deeper into the microscopic origin of
the trend in $T_{c\text{ max}}$ by studying its dependence on the strength of
the Coulomb repulsion, or Hubbard $U$.
Although $U$ and $J$ are directly related in the limit where $U$ is much larger than the bandwidth, for the cuprate compounds $U$ is of the same order as the bandwidth\cite{Werner15Dynamical,jang2016direct}
and hence this relation is not guaranteed to hold. Further on, the value of $U$ is directly influenced 
by the dielectric screening and an investigation of the material dependence of $U$ could therefore be a route to understand the 
correlation between $T_{c\text{ max}}$ and the refractivity sum reported in Ref.~\onlinecite{Mallet13Dielectric}.

We compute the Hubbard $U$ using the constrained 
random-phase approximation (cRPA)\cite{Aryasetiawan04Frequencydependent,Miyake08Screened} as implemented in the FLAPW codes FLEUR and SPEX \cite{Friedrich10Efficient,fleur}.
The cRPA yields both the static (time-averaged) value and the full frequency dependence
of $U$ and allows for a detailed analysis of the screening channels responsible for renormalizing the bare Coulomb
interaction $v$. 

We consider a wide range of hole-doped cuprate compounds starting from the well-studied
La$_2$CuO$_4$ as well as TlBa$_2$CuO$_6$ and HgBa$_2$CuO$_4$ and continuing with the the compounds R(Ba,Sr)$_2$Cu$_3$O$_6$ (R=Y, Yb, Nd, La) that were also studied experimentally in
Ref. \onlinecite{Mallet13Dielectric}. In the latter compounds, changing the ion size yields a change of size of the unit cell and can therefore be 
considered as a change of the "internal pressure" of the compound\cite{Mallet13Dielectric}.  
We also explore the effects of external pressure on $U$ by systematically changing the lattice parameters.

With the exception of La$_2$CuO$_4$, we find a screening-dependent correlation between $T_{c\text{ max}}$ and both the static value of $U$ and the ratio $U/t$,  
suggesting that superconductivity in the cuprates 
is favored by a large onsite Coulomb repulsion. 
Contrary to $J$ we also find that external and internal pressures have the same effect on $U$, that is, $U$ increases with both internal and external pressure.
However, the increase is not sufficiently large to account for the observed increase in $T_{c\text{ max}}$.
In the numerical studies in Refs. \onlinecite{Weber12Scaling} and \onlinecite{Acharya18Metal} it was found that superconductivity was favored by a small charge-transfer energy ($\epsilon_d-\epsilon_p$).
Together with the trend in $U$ in the present paper this suggests that superconductivity may be favored by having a large $U$ and small charge-transfer energy, 
which would lead to a charge-transfer insulating parent compound
 with the lower Hubbard band below the 
$O$ $p$-states. 
Since La$_2$CuO$_4$ both has a large $U$ and a large charge-transfer energy this could offer an explanation of why La$_2$CuO$_4$ does not follow the trend in $U$.

We also consider the full frequency dependent $U(\omega)$ for these compounds. 
We analyze the different screening channels and show that the $p$-$d$ screening channel, that gives rise to peaks around 8-9 eV in all cuprate compounds, is much stronger in La$_2$CuO$_4$ than in the other compounds.
In that sense  La$_2$CuO$_4$ is an unusual case, and may not be a good representative prototype for a general cuprate compound.
Furthermore we show that $U$ is highly material dependent, suggesting that the common assumption of using the same value of $U$ for all cuprate compounds can
yield misleading conclusions.  

\section{Method}
\subsection{cRPA}
\label{Sec:cRPA}
To study the materials dependence of the Hubbard $U$, we use the constrained
random-phase approximation (cRPA) method \cite{Aryasetiawan04Frequencydependent,Miyake08Screened}. In the cRPA method, the screening channels
expressed in terms of the polarizations are decomposed into those within the
model ($P_{d}$) and the rest ($P_{r}$):%

\begin{align}
 P=P_{d}+P_{r}.
\label{crpa1}
\end{align}
It can then be shown that the effective Coulomb interaction among the
electrons residing in the model subspace (the $d$ subspace) is given by%

\begin{align}
U(\omega)=[1-vP_{r}(\omega)]^{-1}v. 
\end{align}
This effective interaction is physically interpreted as the Hubbard $U$, which
is now a function of frequency. This interpretation is based on the fact that
when $U$ is screened by the polarization $P_{d}$ of the model one obtains the
fully screened interaction:%

\begin{align}
W(\omega)=[1-vP(\omega)]^{-1}v=[1-U(\omega)P_{d}(\omega)]^{-1}U(\omega). 
\label{W:crpa}
\end{align}
In practice the polarization is computed from the LDA bandstructure \cite{Kohn65SelfConsistent} within the random phase approximation (RPA)
, which
for a given spin is given by 
\begin{align}
P(\mathbf{r,r}^{\prime };\omega )& =\sum_{\mathbf{k}n}^{\text{occ}}\sum_{\mathbf{k}^{\prime
}n^{\prime }}^{\text{unocc}}\frac{\psi _{\mathbf{k}n}^{\ast }(\mathbf{r})\psi _{\mathbf{k}^{\prime
}n^{\prime }}(\mathbf{r})\psi _{\mathbf{k}^{\prime }n^{\prime }}^{\ast }(\mathbf{r}%
^{\prime })\psi _{\mathbf{k}n}(\mathbf{r}^{\prime })}{\omega -\epsilon _{\mathbf{k}^{\prime
}n^{\prime }}+\epsilon _{\mathbf{k}n}+i\delta }  \notag \\
& -\frac{\psi _{\mathbf{k}n}(\mathbf{r})\psi _{\mathbf{k}^{\prime }n^{\prime }}^{\ast }(%
\mathbf{r})\psi _{\mathbf{k}^{\prime }n^{\prime }}(\mathbf{r}^{\prime })\psi
_{\mathbf{k}n}^{\ast }(\mathbf{r}^{\prime })}{\omega +\epsilon _{\mathbf{k}^{\prime }n^{\prime
}}-\epsilon _{\mathbf{k}n}-i\delta }.  \label{eq:Pol}
\end{align}%

In the LDA the conduction band in the cuprates originates from the antibonding
combination of the Cu $d_{x^2-y^2}$ with the Oxygen $p_x$/$p_y$ orbitals and has $d_{x^2-y^2}$ symmetry. The bonding and 
nonbonding bands, commonly referred to as the Op$_{x/y}$ bands are located around 6 eV below the Fermi energy. Commonly used models 
for the cuprates include 
\begin{itemize}
 \item the one-band model derived from the antibonding conduction band.
 \item the two-band model that apart from the conduction band include the narrow band just below the Fermi energy originating from the  
 hybridization between the Cu 3$d_{z^2}$ and the 
apex Oxygen $p_z$ orbital.
 \item the three-band model that includes the antibonding conduction band as well as the bonding and nonbonding combinations.
 \item the four-band model that include all the above-mentioned bands.
\end{itemize}
In this work we focus on the one-band model since this provides the minimal low-energy model of the cuprates. 

\begin{figure}[t]
  \centering
    \includegraphics[width=0.235\textwidth]{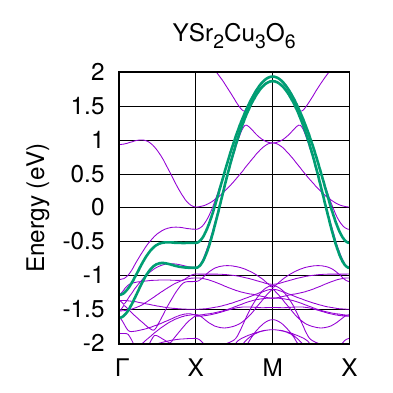}
  \includegraphics[width=0.235\textwidth]{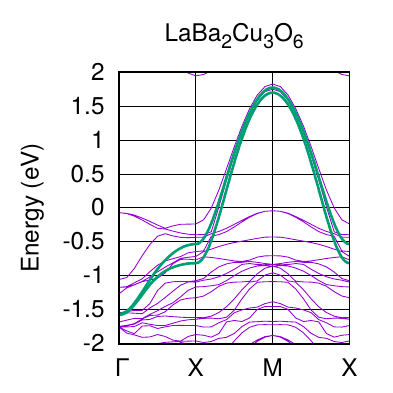}
  \caption{LDA and Wannier interpolated bandstructures of YSr$_2$Cu$_3$O$_6$ and LaBa$_2$Cu$_3$O$_6$. The Fermi energy was set to zero.}
  \label{fig:intband}
\end{figure}

To define the $d_{x^2-y^2}$ model subspace we use Maximally Localised Wannier functions (MLWF:s)
\cite{Marzari97Maximally,Mostofi08wannier90,Sakuma13Symmetryadapted} that are derived from the LDA band structure.
Hence, $P_d$ in Eq.~\ref{crpa1} is the polarization within the $d_{x^2-y^2}$ conduction band. Since the $d_{x^2-y^2}$-band is not isolated we use the disentanglement approach 
\cite{Miyake09Abinitio} to get a well defined one-particle bandstructure and model polarization. In this method the hybridization between the model and the rest is cut
in the Hamiltonian
\begin{equation}
\tilde{H}=\left( 
\begin{array}{cc}
H_{dd} & 0 \\ 
0 & H_{rr} %
\end{array}%
\right).
\label{Eq:Ham}
\end{equation}%
The $r$ subspace polarization is then calculated as $P_{r}=P-P_{d}$, where the
full polarization $P$ and the $d$ subspace polarization $P_{d}$ are calculated 
for the disentangled bandstructure according to Eq. (\ref{eq:Pol}) . In Fig. \ref{fig:intband} we show the Wannier interpolated bandstructures for the 
two compounds YSr$_2$Cu$_3$O$_6$ and LaBa$_2$Cu$_3$O$_6$.

We only compute $U$ for the undoped parent compounds. However, since the metallic screening from within the $d_{x^2-y^2}$ conduction band is removed
within the cRPA, $U$ is expected to be very weakly dependent on the doping. Hence, the values of $U$ for the parent compounds can be used also in the doped cases.

\subsection{Simulation of external pressure}
\label{Sec:pressure}
We consider the effect of external pressure for the compound YbBa$_2$Cu$_3$O$_6$ by scaling the lattice parameters.
With the exception of La$_2$CuO$_4$, where the effect of pressure is approximately isotropic\cite{akhtar1988pressure,Schilling07} (i.e. the $a$ and $c$ lattice parameters are scaled by the same factor)
a generic feature of the cuprates seems to be that the effect of hydrostatic pressure on the $c$ lattice parameter is around twice as big as that of the in-plane lattice parameter $a$\cite{Schilling07}.
Therefore we approximate hydrostatic pressure by scaling $a$ with $x\%$ and $c$ with $2x\%$.
Experimentally $T_{c\text{ max}}$ increases with moderately applied pressure and decreases again at high pressure (larger than 5-7 GPa) \cite{Schilling07,ginsberg1992physical}. 
In Ref. \onlinecite{Armstrong95Crystal} $c$ was shown to decrease about $4\%$ and $a$ about $2\%$ with a pressure of 6 GPa for HgBa$_2$Ca$_2$Cu$_3$0$_{8+\delta}$.
Therefore we consider scalings of $a$ and $c$ below these numbers in this work.

In Ref. \onlinecite{Hardy10Enhancement} both the uniaxial and hydrostatic pressure derivatives of $T_c$ were determined for HgBa$_2$CuO$_{4+\delta}$ close to the optimal doping level.
It was found that $T_c$ increases with a decreasing unit cell area of the Cu-O planes as well as with an increasing separation of the planes.
To investigate the effect of uniaxial pressure on $U$ we also consider scaling only $a$, which corresponds to applying only in-plane pressure.

\section{Computational details}

We use the LDA bandstructure calculated with the
full-potential linearized augmented plane-wave (FLAPW) code FLEUR
\cite{fleur} as a starting point.
The MLWF:s were computed using Wannier90 library \cite{Marzari97Maximally,Mostofi08wannier90,Sakuma13Symmetryadapted} 
and $U$ was computed within the cRPA as implemented in the SPEX-code \cite{Friedrich10Efficient,fleur}.
We only considered spin-polarization for the compounds YbBa$_2$Cu$_3$O$_6$ and 
NdBa$_2$Cu$_3$O$_6$, since the other compounds are not spin-polarized 
within the LDA.
All calculations were converged with respect to the FLAPW basis set, the number of bands used to compute the polarization function, the number of $\mathbf{k}$-points used in 
the LDA as well as cRPA calculation. For example this required the use of between 300-400 bands in the computation of the polarization function for the different compounds.

The bands used to construct the Wannier functions were defined using an energy window, where for each $\mathbf{k}$-point all states with an energy inside the energy window were used 
in the Wannier function construction. In Tab. \ref{tablewindow} we present the energy windows for the different compounds.
For the spin-polarized calculations $U$ was defined as the average matrix element over the two spin channels. However,
both the value of $U$ and the nearest neighbor hopping $t$ were very similar for the two spin-channels. 

For La$_2$CuO$_4$ and TlBa$_2$CuO$_6$ we used the reduced structures in Refs. \onlinecite{La2CuO4CS,TlBa2CuO6CS} while for the remaining materials we use the experimental structures.
The crystal structure for YSr$_2$Cu$_3$O$_6$ was taken from from Ref. \onlinecite{YSr2Cu3O6CS},
YBa$_2$Cu$_3$O$_6$ from Ref. \onlinecite{YBa2Cu3O6CS}, HgBa$_2$CuO$_4$ from Ref. \onlinecite{HgBa2CuO4CS}, YbBa$_2$Cu$_3$O$_6$, NdBa$_2$Cu$_3$O$_6$ and LaBa$_2$Cu$_3$O$_6$ from Ref. \onlinecite{Chu87}.
    
\begin{table}[h]
\caption{Energy windows used in the Wannier function construction (eV).}
\begin{center}
\begin{tabular}{c|c}
LaCuO$_4$	     & -2.5$\rightarrow$2 \\
YSr$_2$Cu$_3$O$_6$   & -2$\rightarrow$2.2   \\
TlBa$_2$CuO$_6$      & -2.2$\rightarrow$3   \\ 
YBa$_2$Cu$_3$O$_6$   & -2$\rightarrow$3   \\
YbBa$_2$Cu$_3$O$_6$  & -3$\rightarrow$2    \\
HgBa$_2$CuO$_4$      &  -2.2$\rightarrow$2 \\
NdBa$_2$Cu$_3$O$_6$  &  -3$\rightarrow$2  \\
LaBa$_2$Cu$_3$O$_6$  & -3$\rightarrow$2 \end{tabular}%
\end{center}
\label{tablewindow}
\end{table}

\begin{figure*}[t]
  \centering
  \includegraphics[width=0.49\textwidth]{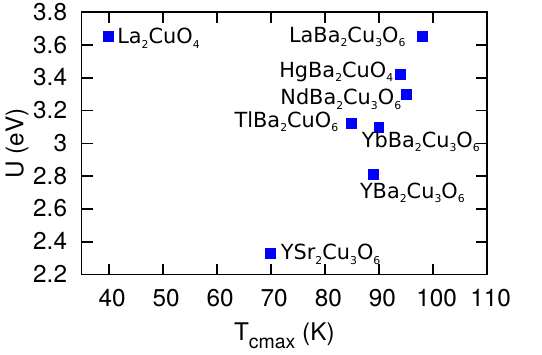}
  \includegraphics[width=0.49\textwidth]{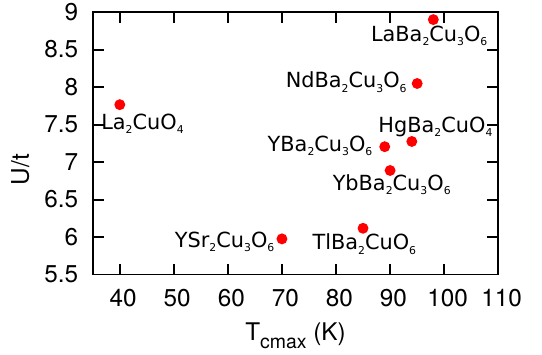}
    \includegraphics[width=0.49\textwidth]{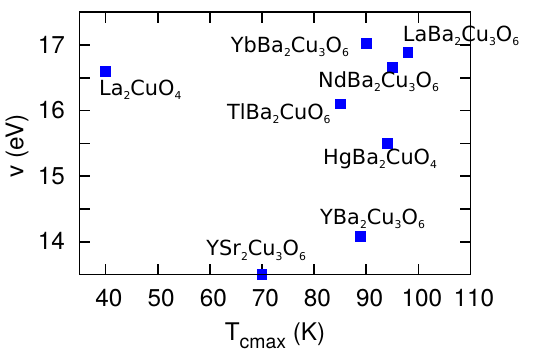}
  \includegraphics[width=0.49\textwidth]{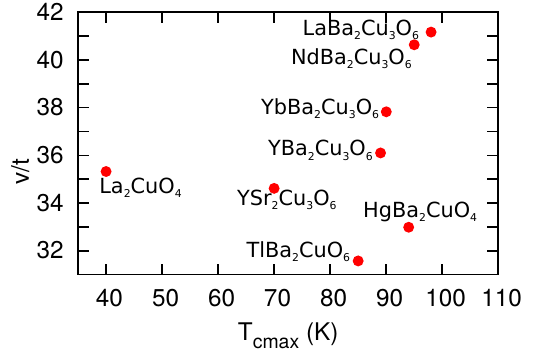}
  \caption{Top left: Static value of $U$ for the parent compounds of a number of hole-doped cuprates in a one-band model. Top right: The ratio $U/t$ for the same compounds. $t$ is the nearest neighbour hopping
  that was derived from a Wannier interpolation. Bottom left: Corresponding matrix element of the bare (unscreened) Coulomb interaction ($v$). Bottom right: The ratio $v/t$.}
  \label{fig:staticU}
\end{figure*}

\section{Results and Discussion}
\subsection{Static Interaction}

In the one-band Hubbard model with a static Coulomb repulsion $U$ and a nearest neighbor hopping $t$ the only
free parameter is the ratio $U/t$. Therefore this ratio provides a good measure of the degree of local correlations.

In Fig.~\ref{fig:staticU} we show the static value of $U$ as well as the ratio $U/t$ for all compounds considered in this work. 
$U/t$ follows an increasing trend with increasing $T_{c\text{ max}}$.
The smallest value of $U/t$ is approximately 6 and the largest approximately 9, which implies a substantial difference of the degree of local correlations
in the compounds.
Since the bandwidths in most of the compounds are similar, $U$ follows the same increasing trend as $U/t$, albeit not as clear.
Hence, the trend in $U/t$ can mainly be attributed to the trend in $U$ and is not an effect of a trend in the hopping parameters.
The only exception to the trend is La$_2$CuO$_4$ which has a remarkably large $U$ compared to the relatively low {\it T}$_c$.
This suggests that La$_2$CuO$_4$, which is typically considered as a prototype of a cuprate high {\it T}$_c$ superconductor, actually is an exceptional case.
Furthermore, assuming that the trend in $U$ implies that high {\it T}$_c$ superconductivity is favored by a large onsite Coulomb repulsion, the fact that
La$_2$CuO$_4$ alludes the trend implies that there are other mechanisms that hamper superconductivity in this compound.
While the compounds R(Ba,Sr)$_2$Cu$_3$O$_6$ all have similar structures with two CuO layers it is interesting to note that both 
TlBa$_2$CuO$_4$ and HgBa$_2$CuO$_4$, which are single layer compounds, also follow the trend. This implies that $U/t$ indeed is an important parameter 
to obtain large {\it T}$_c$ and at least as important as other parameters, such as the number of Cu-O layers. 

The value of $U$ depends both on the screening properties and on the shape and extent of the Wannier basis functions.
The value of the bare interaction $v$, on the other hand, only depends on the shape and extent of the Wannier basis functions. 
More localized Wannier functions yield larger values of $v$.
Since the Wannier functions are derived from the bandstructure, trends in $v$ can be considered as bandstructure effects while
trends in $U$ can depend both on the bandstructure and the screening process. By comparing the values of $v$ (lower left panel) and
$U$ (upper left panel) in Fig.~\ref{fig:staticU} one can conclude that the trend in $U$ is intimately related to the screening in the compounds.
For example TlBa$_2$CuO$_4$ has a larger bare interaction but a smaller value of $U$ than HgBa$_2$CuO$_4$ due to the larger screening in the former compound.
Also in YbBa$_2$Cu$_3$O$_6$, which has the largest value of $v$, the screening is large compared to the other compounds.
In $v/t$ (lower right panel of Fig.~\ref{fig:staticU}) the main exceptions to the trend are HgBa$_2$CuO$_4$ and TlBa$_2$CuO$_4$.

For the compound YbBa$_2$Cu$_3$O$_6$ we simulated the external pressure by scaling the lattice parameters as discussed in Section \ref{Sec:pressure}. 
Our results are summarized in Tab. \ref{table1}.
The effect of external pressure is small ($<2\%$ for reasonable pressures) both on $U$ and $U/t$. Thus $U$ cannot be used to understand the increase of 
$T_{c\text{ max}}$ upon applied external pressure. It is interesting to note that, contrary to the Heisenberg exchange parameter $J$\cite{Mallet13Dielectric}, $U$
follows the same trend upon applied internal and external pressure. However, since also the hopping amplitude increases with a decreasing in-plane Cu distance,  
$U/t$ is unchanged upon applied hydrostatic pressure and decreases with in-plane pressure.

\begin{table}[h]
\caption{Effect of external pressure on $U$ (eV) for YbBa$_2$Cu$_3$O$_6$. External pressure was simulated by reducing the lattice parameters. For in-plane pressure the $a$ lattice parameter 
was reduced by 1$\%$ and for hydrostatic pressure $a$ was reduced by 1$\%$ and $c$ by $2\%$.}
\begin{center}
\begin{tabular}{c|cc}
            & $U$ & $U/t$  \\ \hline
Normal      & 3.1 & 6.9  \\ 
Hydrostatic & 3.2 & 6.9   \\
In-plane    & 3.2 & 6.6   \\
\end{tabular}%
\end{center}
\label{table1}
\end{table}

Taken together our results indicate that $U$ is an important parameter to get high $T_{c\text{ max}}$, and superconductivity is favored by a large onsite Coulomb repulsion.
However, there are exceptions to this trend; La$_2$CuO$_4$ has a relatively large $U$ but low $T_{c\text{ max}}$ and upon applied external pressure the change in $U$ is not sufficient to account for 
the observed increase in $T_{c\text{ max}}$. This indicates that $U$ or $U/t$ are not the only important parameter for high $T_c$ superconductivity in the cuprates, rather
there seem to be many competing mechanisms that taken together determine whether a material has a high $T_c$ or not. 
The shape of the Fermi surface, as indicated by the trend in $T_{c\text{ max}}$ with $t'/t$ in Ref. \onlinecite{Pavarini01Band}, is an example of one such important parameter.  
\begin{figure}[t]
  \centering
  \includegraphics[width=0.49\textwidth]{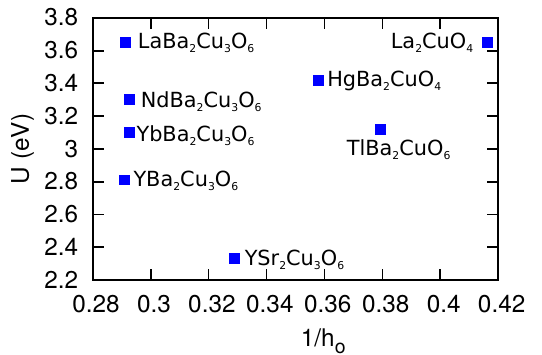}
  \caption{Static value of $U$ as a function of the average inverse apical Oxygen distance $1/h_O = \frac{1}{\sqrt{2}}(1/h_{O_1} + 1/h_{O_2})$, 
 where $h_{O_{1/2}}$ is the distance to the apical Oxygen above/below the Cu ion in the CuO plane.}
  \label{fig:UvsOD}
\end{figure}

\subsection{Comparison to other calculations}
In this section we compare our results to other similar studies. The two main studies we will focus on are the ones by Jang {\it et al}. 
\cite{jang2016direct} that compared $U$ for a number electron and hole-doped cuprates as well as Hirayama {\it et al}.\cite{Hirayama18Ab} who derived the low-energy
Hamiltonian for La$_2$CuO$_4$ as well as HgBa$_2$CuO$_4$ in the one, two and three-band models using the MACE-scheme\cite{Imada10Electronic,Hirayama13Derivation,Hirayama17Low}.
We also briefly discuss the numerical studies by Weber {\it et al}.\cite{Weber12Scaling} and Acharya {\it et al}.\cite{Acharya18Metal}, using cluster DMFT and $GW$+DMFT based schemes respectively, showing a correlation
between the charge-transfer energy and $T_c$, which indicates that superconductivity is favored by a small charge-transfer gap.

\subsubsection{Comparison to Jang et al.} 
In Ref. \onlinecite{jang2016direct} Jang {\it et al}. computed the static value of $U$ using the cRPA for the parent compounds of a number of both hole- and electron-doped cuprates.
Of specific interest to this work they computed $U$ for La$_2$CuO$_4$ and HgBa$_2$CuO$_4$. For both these compounds they obtained a value of $U$ which is substantially smaller
than the ones in this work (3.15 eV for La$_2$CuO$_4$ compared to 3.65 eV in this work and 2.15 eV for HgBa$_2$CuO$_4$ compared to 3.42 eV in this work).
The origin of this difference is the difference in methods when computing the model polarization. We used the disentanglement approach\cite{Miyake09Abinitio} described in Section \ref{Sec:cRPA}
while Ref. \onlinecite{jang2016direct} used a weighting approach\cite{Sasiuglu11Effective} where the polarization is computed for the original LDA bandstructure and $P_d$ is defined as
\begin{align}
&P_{d}(\mathbf{r,r}^{\prime };\omega ) = \sum_{\mathbf{k}n}^{\text{occ}}\sum_{\mathbf{k}^{\prime
}n^{\prime }}^{\text{unocc}} \Bigl( \frac{\phi _{\mathbf{k}n}^{\ast }(\mathbf{r})\phi _{\mathbf{k}^{\prime
}n^{\prime }}(\mathbf{r})\phi _{\mathbf{k}^{\prime }n^{\prime }}^{\ast }(\mathbf{r}%
^{\prime })\phi _{\mathbf{k}n}(\mathbf{r}^{\prime })}{\omega -\epsilon _{\mathbf{k}^{\prime
}n^{\prime }}+\epsilon _{\mathbf{k}n}+i\delta }  \notag \\
 &-\frac{\phi _{\mathbf{k}n}(\mathbf{r})\phi _{\mathbf{k}^{\prime }n^{\prime }}^{\ast }(%
\mathbf{r})\phi _{\mathbf{k}^{\prime }n^{\prime }}(\mathbf{r}^{\prime })\phi
_{\mathbf{k}n}^{\ast }(\mathbf{r}^{\prime })}{\omega +\epsilon _{\mathbf{k}^{\prime }n^{\prime
}}-\epsilon _{\mathbf{k}n}-i\delta } \Bigr) P_{\mathbf{k}n}P_{\mathbf{k}'n'}.  \label{eq:cRPAweighting}
\end{align}
$P_{\mathbf{k}n}$ is the probability that the electron in state $|\phi _{\mathbf{k}n}\rangle$ resides in the $d$-subspace.
This method generally yields smaller values of $U$ since not all metallic screening from the correlated (disentangled) band is removed.
Furthermore, the final aim of our calculations is to use the $U$ values together with a Hamiltonian or hopping parameters in e.g. LDA+DMFT or
$GW$+EDMFT calculations. The Hamiltonian for this type of calculation would exactly correspond to the $d$-block of the disentangled Hamiltonian in Eq. \ref{Eq:Ham}.
Hence, in the disentanglement approach
both the $U$-matrix and the hopping parameters are derived from the disentangled bandstructure, which is not the case in the weighting approach. 
We therefore consider the disentanglement approach to be a more appropriate method
in this case. 

In Ref. \onlinecite{jang2016direct} the focus was on the comparison between electron-doped and hole-doped cuprates and a general tendency of electron-doped cuprates to have a smaller $U$ was found.
However, also some of the hole-doped cuprates had similarly small values of $U$ which lead the authors to conclude that
the strong correlation enough to induce Mott gap may not be a prerequisite for the high-$T_c$ superconductivity.  
In this work we consider a wider range of hole-doped cuprates. Even though $U$ is generally larger using the disentanglement approach
($U$ for HgBa$_2$CuO$_4$ using the disentanglement approach is still larger than the value of $U$
computed for La$_2$CuO$_4$ with the weighting approach in Ref. \onlinecite{jang2016direct})
at first sight our results seems 
to strengthen the conclusions in Ref. \onlinecite{jang2016direct}. $U$ for YSr$_2$Cu$_3$O$_6$ for example is only 2.33 eV which is approximately 0.6 times the bandwidth, and hence 
cannot be considered to be deep in the Mott-insulating regime. 
However, as discussed in e.g. Ref. \onlinecite{Werner15Dynamical}, cRPA for a pure one-band model could potentially underestimate $U$ due to the large spread of the Wannier basis states
in this model.
Hence, one should mainly focus on the trend rather than 
 the absolute values of $U$ in a one-band model.
 The trend in the static value implies that,
even though strong correlation enough to induce Mott gap may or may not be a prerequisite for the high-$T_c$ superconductivity, superconductivity is favored by 
strong local Coulomb repulsions.  
 
 Jang {\it et al}. found that the electron-doped cuprates, which have smaller $T_{c\text{ max}}$ than their hole-doped counterparts, also had smaller values of $U$.
 This observation fits with the trend reported in this paper. However, for Hg-based compounds with different number of CuO-layers, they also found that 
 $U$ for the triple layer compound HgBa$_2$Ca$_2$Cu$_3$O$_8$ is smaller than the corresponding value of the double and single layer compounds, even though
 $T_{c\text{ max}}$ increases with the number of layers.
 This result contradicts the trend and therefore further illuminates the complexity of the problem with many competing mechanisms.
 In this particular case it points to two competing mechanisms to achieve high {\it T}$_c$ superconductivity, namely a large $U$ on one hand and many CuO layers on the other hand.

 Another interesting observation in Ref. \onlinecite{jang2016direct} was a correlation between $U$ and the average inverse apical Oxygen distance $1/h_O = \frac{1}{\sqrt{2}}(1/h_{O_1} + 1/h_{O_2})$, 
 where $h_{O_{1/2}}$ is the distance to the apical Oxygen above/below the Cu ion in the CuO plane.
 From Fig. \ref{fig:UvsOD} it is clear that, while we reproduce this correlation for YSr$_2$Cu$_3$O$_6$, TlBa$_2$CuO$_6$, HgBa$_2$CuO$_4$ and La$_2$CuO$_4$, the remaining compounds considered in this 
 work do not show any correlation between $U$ and $1/h_O$. For these materials the screening from the charge reservoir layers are important as will be discussed more in detail below.

\begin{figure*}[t]
  \centering
  \includegraphics[width=0.49\textwidth]{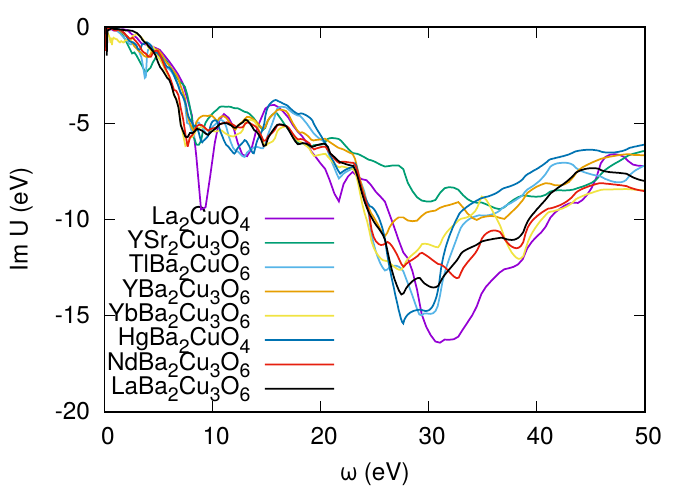}
    \includegraphics[width=0.49\textwidth]{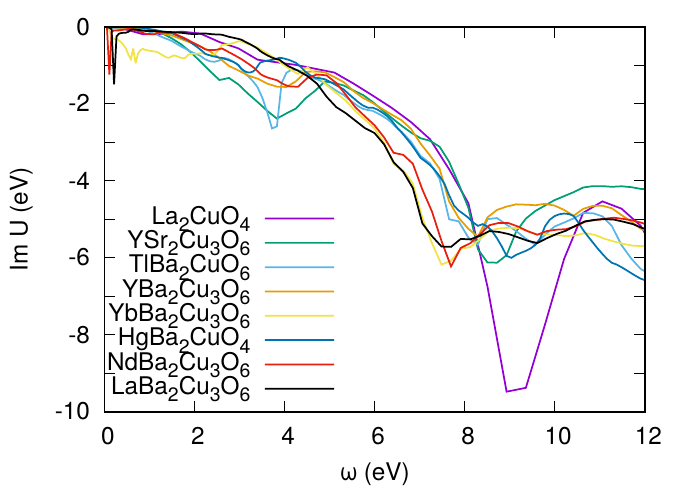}
  \caption{Imaginary part of the frequency dependent $U$ for the different compounds. The right panel shows a close-up on the low-frequency structure.}
  \label{fig:freqdep}
\end{figure*}
 
\subsubsection{Comparison to Hirayama et al.} 
Hirayama {\it et al}.\cite{Hirayama18Ab} calculated the effective low-energy Hamiltonians for La$_2$CuO$_4$ as well as HgBa$_2$CuO$_4$ in the one, two and three band models using the MACE scheme\cite{Imada10Electronic,Hirayama13Derivation,Hirayama17Low}.
Of interest to this work are their values of $U$ in the one-band model (5.00 eV for La$_2$CuO$_4$ and 4.37 eV for HgBa$_2$CuO$_4$) which are substantially larger than the ones in this work.
The MACE-scheme involves a cRPA calculation using the disentanglement approach, but for the 17-bands closest to the Fermi-energy (which includes the Cu $d$-bands and Oxygen $p$-bands) they
computed the polarization from the $GW$ quasiparticle bandstructure rather than the LDA. This yielded smaller screening and hence larger values of $U$. 

The reason that La$_2$CuO$_4$ has a relatively large $U$ was also analyzed.
It was concluded that La$_2$CuO$_4$ has a larger value of $U$ than HgBa$_2$CuO$_4$ because the Oxygen $p$ orbitals are farther below the
Cu $d_{x^2-y^2}$ orbitals in La$_2$CuO$_4$ which yields a different (more localized) character of the antibonding conduction band. This analysis would imply that both $U$ and the bare Coulomb interaction $v$
should be larger for La$_2$CuO$_4$, which also agrees with our results in Fig.~\ref{fig:staticU}.

\subsubsection{Comparison to Weber et al. and Acharya et al.}
In Ref. \onlinecite{Weber12Scaling} Weber {\it et al}. studied {\it T}$_c$ as a function of the charge-transfer energy $(\epsilon_d-\epsilon_p)$ as well as the hopping parameters 
using cluster DMFT for the three-band model with fixed value of $U_{dd}=$ 8 eV. It was found that the charge-transfer energy shows an antilinear correlation with the static order parameter, i.e. decreasing $(\epsilon_d-\epsilon_p)$
yielded a larger superconducting order parameter. It was also shown that the charge-transfer energy, computed from the LDA bandstructure, displayed an antilinear correlation with the experimental {\it T}$_{c\text{ max}}$
for a large number of cuprates. 

In Ref. \onlinecite{Acharya18Metal} Acharya {\it et al}. studied how the displacement of the apical Oxygen in La$_2$CuO$_4$ affects the superconducting order parameter,
optical gap as well as spin and charge susceptibilities using a one-shot combination of quasiparticle self-consistent $GW$ (QPSC$GW$) and DMFT. They used a 
static $U$ of 10 eV which is substantially larger than 
the static cRPA value\cite{Werner15Dynamical}, but ignored the frequency dependence. This large $U$ value was motivated by comparing full and restricted QPSC$GW$ calculations in Ref. \onlinecite{choi2016first}.
It was found that pristine LaCuO$_4$ was Mott-insulating but increasing the distance between the apical Oxygen and the Cu-O plane ($\delta$) yields a cross-over to a charge-transfer insulator (CTI). 
Increasing $\delta$ further shrinks the CTI gap and the gap collapses at the critical value $\delta_c = 0.045$Å. They estimated $T_{c\text{ max}}$ from their calculated values of the 
superconducting order parameter and found that {\it T}$_{c\text{ max}}$ increased with increasing $\delta$ until it reached its maximum value at $\delta=\delta_c$.
These results support the conclusions by Weber \emph{et al.} that superconductivity is favored by a small charge-transfer gap.

Since an estimation of the charge-transfer gap requires the use of a three-band model, straightforward comparisons between these two works and our results are difficult.
However, as discussed in detail in Ref. \onlinecite{Werner15Dynamical}, in practical calculations for the three-band model using e.g. LDA+DMFT, $U_{pp}$ and $U_{pd}$
are typically ignored. In such, so called $d$-$dp$ model calculations, the $p$-$d$ screening should be included in the effective $U_{dd}$. Hence, the only difference between the $U_{dd}$ in the one-band model and 
the effective $U_{dd}$ in the $d$-$dp$ three-band model comes from the Wannier basis functions, which are more localized in the latter case. Therefore it is reasonable to assume that the 
effective $U_{dd}$ in the three-band model will follow the same trend as in the one-band model but with larger overall values. 
Both Weber {\it et al}. and Acharya {\it et al}. effectively decreased the charge-transfer energy while keeping $U_{dd}$ fixed. This yields a transition from a Mott insulator to a pure charge-transfer insulator with 
the lower Hubbard band below the Oxygen $p$-states.
The same effect can be reached by keeping the charge-transfer gap fixed and 
increasing the $U$. The results in this work combined with the studies above therefore suggest that superconductivity in the hole-doped cuprates is favored by 
having a CTI parent compound.This can be achieved by having a large
$U$ and/or a small charge-transfer energy.
Due to the large charge-transfer energy in La$_2$CuO$_4$  the lower Hubbard band is in the same energy range as the Oxygen $p$ bands\cite{Werner15Dynamical} in spite of the large $U$-value, 
which can explain why La$_2$CuO$_4$ does not follow the trend in Fig. \ref{fig:staticU}.

\subsection{Frequency dependence}
In Fig.~\ref{fig:freqdep} we show the imaginary part of the frequency dependent $U$ for a selected number of compounds. 
The main features that can be observed in all materials is a subplasmonic peak around 8-9 eV,
as well as the main bulk plasmon around 30 eV.
The 8-9 eV peak originates from screening from the Oxygen $p$ bands below the Fermi energy\cite{Werner15Dynamical}.
To provide a rough estimation of the position of the peak we will consider a two-level system.
The poles of the response function for a two-level system is given by
\begin{equation}
\Omega _{nn^{\prime }}=\sqrt{\Delta \epsilon _{nn^{\prime
}}^{2}+2J_{nn^{\prime }}\Delta \epsilon _{nn^{\prime }}},
\end{equation}%
where $\Delta \epsilon _{nn^{\prime }}$ is the energy difference between the
states and $J_{nn^{\prime }}$ the exchange interaction between
the states.
For La$_2$CuO$_4$ the Oxygen $p$ bands are relatively far below the Fermi energy which implies that $\Delta \epsilon _{nn^{\prime }}$ is large, 
and therefore the $p$-$d$ peak appears at relatively
high energy in $U$.

It is also interesting to note that the $p$-$d$ peak is much more pronounced in La$_2$CuO$_4$ than in the other compounds. 
If the $p$-$d$ screening acts destructively for superconductivity this could offer an alternative explanation why La$_2$CuO$_4$ eludes the trend in the static $U$ in Fig. \ref{fig:staticU}.
However, it is possible that the large onsite Coulomb interaction, which is not accounted for when computing the cRPA $U$, suppresses the $p$-$d$ screening channel in the real material.
The tendency of the cRPA to overestimate the low-energy screening channels between narrow bands close to the Fermi energy has also been indicated in model studies in Refs. 
\onlinecite{Shinaoka15Accuracy,han2018investigation,honerkamp2018limitations}.

\begin{figure*}[ht]
  \centering
  \includegraphics[width=0.49\textwidth]{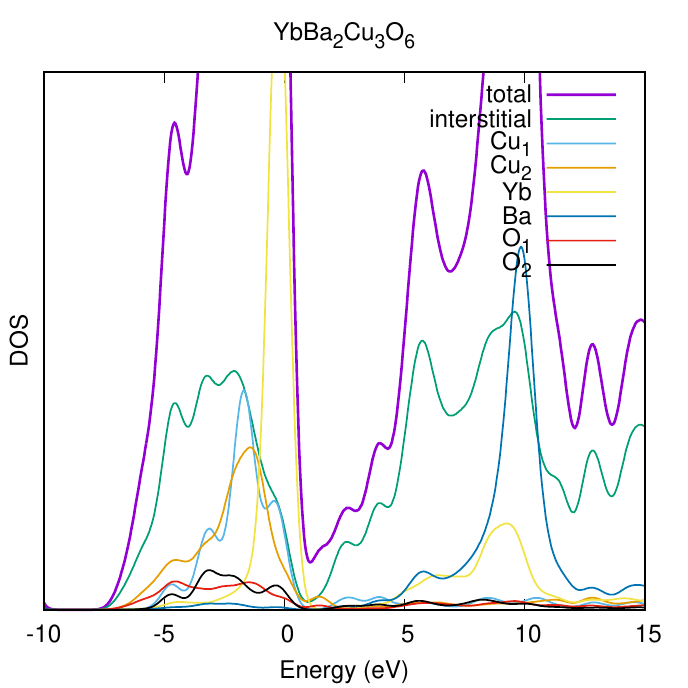}
  \includegraphics[width=0.49\textwidth]{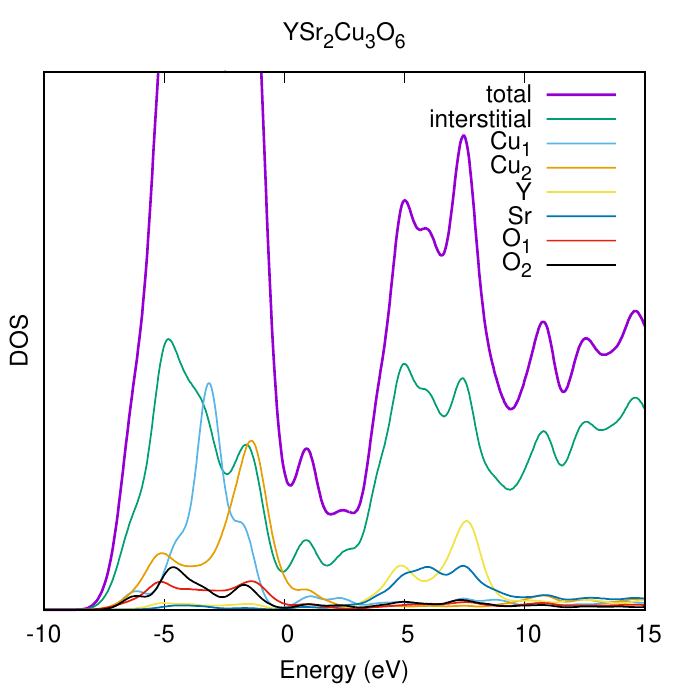}
  \caption{Density of states (DOS) for YSr$_2$Cu$_3$O$_6$ (left) and YbBa$_2$Cu$_3$O$_6$ (right). The partial weight in the different muffin-tin regions are also shown.
  Here Cu$_2$ is the Cu ion in the CuO plane, O$_2$ is the apical Oxygen and O$_1$ the in-plane Oxygen. The Fermi energy was set to zero.}
  \label{fig:dos}
\end{figure*}

\begin{figure}[t]
  \centering
  \includegraphics[width=0.49\textwidth]{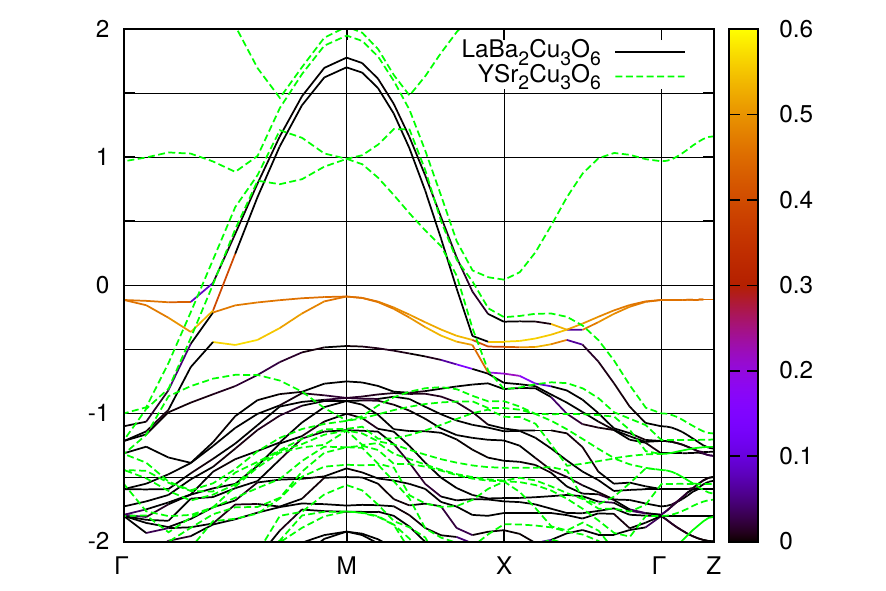}
  \caption{LDA bandstructure of YSr$_2$Cu$_3$O$_6$ and LaBa$_2$Cu$_3$O$_6$. For LaBa$_2$Cu$_3$O$_6$ we constructed Maximally Localized Wannier Functions for the entire isolated
   set of 33 bands around the Fermi energy. The color-coding shows the projection onto the Wannier functions centered on the out-of-plane Cu of $d_{xz}$ and $d_{yz}$ symmetries.
   The Fermi energy was set to zero.}
  \label{fig:band}
\end{figure}

In addition to these features the compounds with a rare-earth element RBa$_2$Cu$_3$O$_6$ (R=Yb,Nd,La) exhibit a well pronounced low frequency structure around 0.5-1eV.
In Fig. \ref{fig:dos} we compare the DOS of YSr$_2$Cu$_3$O$_6$ where this peak is absent and YbBa$_2$Cu$_3$O$_6$ which displays the low-energy peak in Im$U$. 
From this comparison it is tempting to conclude that the metallic screening originates from the Yb spectral weight close to the Fermi energy, which originates 
from the narrow $4f$ band in the LDA bandstructure. However, these states are highly localized on the rare-earth ion and do not contribute much to the screening
on the Cu ion. This is also apparent since LaBa$_2$Cu$_3$O$_6$ does not have these 4$f$ bands but yet display a strong metallic screening.
Rather, the origin of the low-energy screening channel is apparent in the bandstructure. In Fig. \ref{fig:band} the bandstructure of LaBa$_2$Cu$_3$O$_6$ and 
YSr$_2$Cu$_3$O$_6$ are compared. LaBa$_2$Cu$_3$O$_6$ exhibits a narrow band close to the Fermi energy, 
which can be deduced to originate from the $d_{xz}$ and $d_{yz}$ orbitals from the out of plane Cu ion. 
The closeness to the Fermi-energy and the strong hybridization gives rise to the strong screening channel 
in Im$U$.
In YSr$_2$Cu$_3$O$_6$, this band is at much lower energy and not as strongly hybridized with the Cu$d_{x^2-y^2}$ conduction band.

\subsection{Screening analysis}

By keeping the basis functions fixed and making use of an energy window to selectively remove different screening channels it is possible to dissect which screening channels that
contribute to the trend.
Screening from all states within the energy window as well as screening due to transitions from states within the energy window to the model subspace ($d_{x^2-y^2}$ conduction band) is removed.
The bare interaction then corresponds to the case with an infinite energy window.
We consider the following windows:
\begin{enumerate}
 \item -8$\rightarrow$2 eV
 \item -8$\rightarrow$12 eV
\end{enumerate}
Window 1 excludes the $p$-$d$ screening as well as the additional low-energy screening channels while Window 2 also excludes the screening
to higher lying bands.

\begin{figure}[t]
  \centering
  \includegraphics[width=0.49\textwidth]{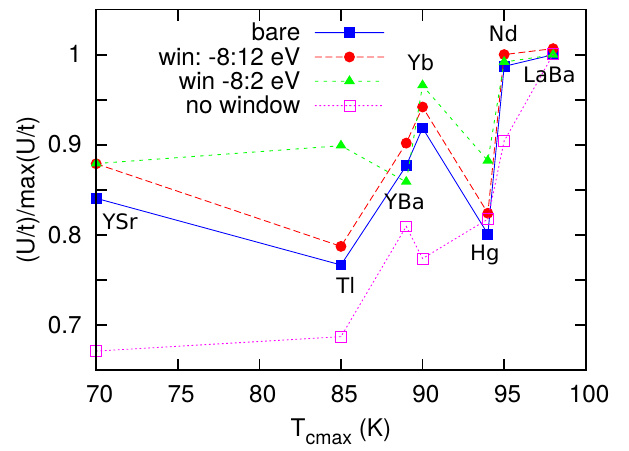}
  \caption{In this figure we used an additional external energy window to selectively remove different screening channels in the polarization. Screening from all states within the energy window as well as screening due to transitions from states within the energy window to the model subspace ($d_{x^2-y^2}$ conduction band) is removed.
  $U$ corresponds to the interaction without an additional energy window and the bare interaction the interaction with an infinite energy window. 
  Each data point corresponds to the ratio $U/t$ for one material with one energy window. The different values that correspond to the same energy window are connected.
  For a better comparison we scaled the ratio $U/t$ by its maximum value for each energy window.
  The order of the compounds are YSr$_2$Cu$_3$O$_6$ $\rightarrow$ TlBa$_2$CuO$_6$ $\rightarrow$ YBa$_2$Cu$_3$O$_6$ $\rightarrow$ YbBa$_2$Cu$_3$O$_6$ $\rightarrow$
  HgBa$_2$CuO$_4$ $\rightarrow$ NdBa$_2$Cu$_3$O$_6$ $\rightarrow$ LaBa$_2$Cu$_3$O$_6$.}
  \label{fig:windows}
\end{figure}

Since we are not interested in the absolute values but only the relative trend we show the value of $U/t$ scaled by its maximum value for each energy window in Fig.~\ref{fig:windows}.
For the large energy window (-8$\rightarrow$12 eV) the picture is almost identical to the case with the bare interaction. This
shows that the high energy screening affects all compounds in the same way and that the material specific screening is related to the low-energy screening channels within the energy window.
This can be understood from the DOS in Fig.~\ref{fig:dos}. For energies higher than 12 eV the majority of the spectral weight comes from the interstitial region of the FLAPW basis set and 
therefore corresponds to broad bands that are not expected to yield a very material specific screening.

The inclusion of the screening to states between 2 to 12 eV (see window -8$\rightarrow$2 eV) is dramatic and highly material specific. This is expected since this energy region contains a 
large spectral weight on the atoms in the charge reservoir layers (Sr/Ba, Y etc.), which is highly material specific. 
However, it is only upon the inclusion of the additional low energy screening channels in $U$ that a clear trend can be observed. 
Hence the trend cannot be attributed to any specific screening channel but all low energy screening channels, within the -8 $\rightarrow$ 12 eV energy window collectively
contribute to the trend.

\section{Conclusions}
We have computed the effective Coulomb interaction $U$ for the one-band model for the parent compounds of a number of hole-doped cuprate superconductors using the constrained random-phase approximation (cRPA).
We find a screening dependent trend between the maximum superconducting transition temperature ($T_{c\text{ max}}$) and the static screened interaction $U$, 
suggesting that superconductivity is favored by a large onsite effective Coulomb interaction.
The only exception to the trend is La$_2$CuO$_4$ which has a relatively large value of $U$ but the smallest $T_{c\text{ max}}$.
From our data we suggest that $U$ is one out of many competing parameters to achieve high {\it T}$_c$ superconductivity and that there are other mechanisms that hamper superconductivity in
La$_2$CuO$_4$, such as the large charge-transfer energy\cite{Weber12Scaling,Acharya18Metal}. We also study the frequency dependence of $U$ and explain the different features.
One of the most dominant features, present in all the studied compounds, is a peak in Im$U(\omega)$ at around 8-9 eV, which originates from screening from the O$p_{x/y}$ bands to the 
$d_{x^2-y^2}$ conduction band ($p$-$d$ screening). This peak is much more pronounced in La$_2$CuO$_4$ than in the other compounds which leads us to suggest that, apart from the large charge-transfer energy, the strong $p$-$d$ screening 
in La$_2$CuO$_4$ could be one possible mechanism that hamper superconductivity in this compound. For the compounds RBa$_2$Cu$_3$O$_6$, R=La,Nd,Yb, we find an additional
low-energy screening channel due to screening from the band derived from the out of plane Cu $d_{xz}$ and $d_{yz}$ states. 
This band is close to the Fermi energy and strongly hybridized with the $d_{x^2-y^2}$ conduction band for these compounds which yields an unusually strong screening mode.

\begin{acknowledgments}
This work was supported by
the Swedish Research Council. The computations were performed on resources provided by the Swedish National Infrastructure for Computing (SNIC) at LUNARC
We would like to thank O.K. Andersen for useful inputs and discussions.
We would also like to thank C. Friedrich and S.
Bl\"{u}gel for providing us with their FLAPW code.
\end{acknowledgments}

%

\end{document}